\documentclass[12pt]{article}
\usepackage[dvips]{epsfig}

\begin{document}

\title{Flat cosmological models with massive scalar field in gauge theories of gravity}

\author{G.V. Vereshchagin \\
{\it \small Belorussian State University, Theoretical Physics Department,} \\
{\it \small Skorina ave.  4, 220050, Minsk, Republic of Belarus}}

\date{}

\maketitle

\begin{abstract}
Solutions of gravitational equations of gauge theories of gravity in homogeneous isotropic world with massive scalar field are investigated in the case of flat cosmological models. Special attention is dedicated to general behavior of solutions on contraction stage. It is shown, that on expansion stage inflationary solutions are generic feature of the model. At the same time on contraction stage an effective equation of state is similar to the case of massless scalar field $p=\rho$ at the beginning of evolution and tends to the equation of ultrarelativistic gas $p=\rho/3$. The Hubble parameter tends to some negative value on contraction stage, depending on the mass of the scalar field. Nonsingular solutions in this model are unstable.
\end{abstract}

\section{Introduction}

Inflationary cosmology \cite{AS}-\cite{Linde} allows to resolve a number of puzzles of Big Bang model. It became a standard paradigm since observational data favor a flat cosmological model and a scale-invariant spectrum of initial perturbations, that are main predictions of inflation \cite{Linde}.

Cosmological singularity is a natural attribute of inflationary models in General Relativity (GR). In spite of the fact, that nonsingular inflationary solutions exist in the case of closed models \cite{Sta78},\cite{BGZH}, they cannot be used in order to construct viable nonsingular cosmological models. Their measure is equal to zero, and, moreover, on the contraction stage such solutions are unstable and have the effective equation of state $p=\rho$ \cite{BGZH}.

When departure from a Riemannian spacetime is allowed, the geometry becomes more general, such as affine geometry with torsion and nonmetricity in addition to curvature tensor. A review of gauge theories of gravitation (GTG) in such spacetimes can be found in \cite{Hehl}. This could happen for example during the inflationary stage where in addition to the metric tensor a scalar dilaton \cite{inf} induced by a Weyl geometry appears.

One of the most important feature of GTG is a possibility to avoid cosmological singularity \cite{phlett80}. Indeed, nonsingular cosmological models were built on the basis of GTG \cite{minkd85}. In particular, inflationary nonsingular cosmological solutions were obtained \cite{mv1}.

However, the question on the generality of inflationary and (or) nonsingular solutions were not adequately explored within the framework of GTG in comparison with the case of GR (see \cite{KLM} for a review). In this paper solutions of cosmological equations of GTG in the case of flat models in the Universe, filled by massive scalar field are analyzed and discussed. General behavior of such solutions is investigated both on contraction and expansion stages by means of qualitative theory of dynamical systems.

Cosmological equations of GTG contain indefinite parameter $\beta$ (see below), that has a dimension of inverse energy density and determines roughly the scale, when predictions of GTG diverge from that of GR. Notice, that regular cosmological models \cite{mv1} were obtained in the case $\beta<0$.

The paper is organized as follows. At the section 2 equations for a dynamical system are derived. In section 3 qualitative analysis is carried out. In section 4 the phase space at infinity is analyzed. Section 5 contains discussion. Conclusions follows in the last section.

\section{Basic equations}

Homogeneous isotropic models in GTG are described cosmological equations \cite{phlett80}

\begin{equation}
\label{gcfe}
\frac{k}{R^2}+\left\{\frac{d}{dt}\ln\left[R\sqrt{\left|1-\beta\left(\rho-
3p\right)\right|}\,\right]\right\}^2=\frac{8\pi }{3M_{p}^{2}}\,\frac{\rho-
\frac{\beta}{4}\left(\rho-3p\right)^2}{1-\beta\left(\rho-3p\right)}
\, ,
\end{equation}

\begin{equation}
\label{Heqn1}
\frac{\left[\dot{R}+R\left(\ln\sqrt{\left|1-\beta\left(\rho-
3p\right)\right|}\,\right)^{\cdot}\right]^\cdot}{R}= -\frac{8\pi
}{6M_{p}^{2}}\,\frac{\rho+3p+\frac{\beta}{2}\left(\rho-3p\right)^2}{
1-\beta\left(\rho-3p\right)}\, .
\end{equation}
where $R(t)$ is the scale factor in the Robertson-Walker metric,
$k=-1,0,+1$ for open, flat and closed models respectively, $M_{p}$ is
planckian mass, $\rho(t)$ and $p(t)$ are energy density and pressure, $\beta$ is indefinite parameter, and a dot denotes the differentiation with respect to time. \footnote{Hereinafter the system of units with $\hbar=c=1$ is used.}

Besides equations (\ref{gcfe}-\ref{Heqn1}) gravitational equations of GTG lead to the following relation for torsion function $S$ and nonmetricity function
$Q$

\begin{equation}
\label{SQ}
S-\frac{1}{4}Q=-\frac{1}{4}\frac{d}{dt}\ln[1-\beta(\rho-3p)].
\end{equation}

Note, that in Poincare gauge theory of gravity $Q=0$ and equation (\ref{SQ}) determines the torsion function.  In metric-affine theory of gravity equation (\ref{SQ}) may describe three kinds of models: in the Riemann-Cartan spacetime $Q=0$, in the Weyl spacetime $S=0$, in the Weyl-Cartan spacetime ($S\neq 0, Q\neq 0$), the function $S$ is proportional to the function $Q$ \cite{garCQG}.

The conservation law has the usual form

\begin{equation}
\label{cl}
\dot{\rho}+3H(\rho+p)=0,
\,\,\,\,\,\,\,\,\,\,\,\,\,\,\,\,\,\,\,\,\,\,\,
H(t)\equiv\frac{\dot{R}}{R},
\end{equation}
where $H(t)$ is a Hubble parameter.

In the case of massive scalar field minimally coupled to gravity the energy density and pressure are

\begin{equation}
\begin{array}{rcl}
\label{phicon}
\rho=\frac{1}{2}\dot{\varphi}^2+\frac{1}{2}m^2\varphi^2,
\,\,\,\,\,\,\,\,\,\,\,\,\,\,\,\,\,\,\,\,\,\,\,
p=\frac{1}{2}\dot{\varphi}^2-\frac{1}{2}m^2\varphi^2,
\end{array}
\end{equation}
where $m$ is the mass of the scalar field $\varphi(t)$.

Then taking into account (\ref{cl}-\ref{phicon}) the cosmological equations (\ref{gcfe}-\ref{Heqn1}) can be rewritten in the following way

\begin{equation}
\label{gcfe2}
\begin{array}{l}
\begin{displaystyle}
\frac{k}{R^2}+\frac{H^2}{Z^2}\left[1-2\beta(\dot\varphi^2+m^2\varphi^2)\right]^2-\frac{H}{Z^2}6m^2\beta \varphi \dot\varphi\left[1-2\beta(\dot\varphi^2+m^2\varphi^2)\right]+
\end{displaystyle} \\
\begin{displaystyle}
\;\;\;\;\;\;\;\;\;\;\;\;\;\;\;\;\;\;
\;\;\;\;\;\;\;\;\;\;\;\;\;\;\;\;\;\;
+\frac{1}{Z^2}9m^4\beta^2\dot\varphi^2\varphi^2=
\frac{4\pi }{3M_{p}^2 Z}\,
\left[\dot{\varphi}^2+m^2\varphi^2-\frac{\beta}{2}\left(\dot{\varphi}^2-2m^2\varphi^2\right)^2\right],
\end{displaystyle}
\end{array}
\end{equation}

\vskip4mm

\begin{equation}
\label{Heqn2}
\begin{array}{l}
\begin{displaystyle}
\frac{\dot{H}}{Z}\left[1-2\beta(\dot\varphi^2+m^2\varphi^2)\right]+\frac{H^2}{Z^2}\left[(1-2\beta m^2\varphi^2)^2-2\beta^2\dot\varphi^4+17\beta\dot\varphi^2(1-2\beta m^2\varphi^2)\right]+
\end{displaystyle} \\
\begin{displaystyle}
\;\;\;\;\;\;\;\;\;\;\;\;\;
+\frac{H}{Z^2}12m^2\beta \varphi \dot\varphi\left[1-2\beta(\dot\varphi^2+m^2\varphi^2)\right]-
\end{displaystyle} \\
\begin{displaystyle}
\;\;\;\;\;\;\;\;\;\;\;\;\;\;
\;\;\;\;\;\;\;\;\;\;\;\;\;\;
-\frac{3\beta m^2}{Z^2}\left[2\beta m^4\varphi^4+\dot\varphi^2(\beta \dot\varphi^2+1)+m^2\varphi^2(3\beta \dot\varphi^2-1)\right]=
\end{displaystyle} \\
\begin{displaystyle}
\;\;\;\;\;\;\;\;\;\;\;\;\;\;\;\;\;\;\;\;
\;\;\;\;\;\;\;\;\;\;\;\;\;\;\;\;\;\;\;\;
=\frac{4\pi}{3M_{p}^{2}}\left[m^2\varphi^2-2\dot{\varphi}^2-\frac{\beta}{2}\left(2m^2\varphi^2
-\dot{\varphi}^2\right)^2\right],
\end{displaystyle}
\end{array}
\end{equation}

\vskip4mm

where $Z=1+\beta\left[\dot{\varphi}^2-2m^2\varphi^2\right]$.

The conservation law (\ref{cl}) in the case under consideration is reduced to the scalar field equation
\begin{equation}
\label{scl}
\begin{array}{rcl}
\ddot{\varphi}+3H\dot \varphi+m^2\varphi=0.
\end{array}
\end{equation}

In the case of flat cosmological models ($k=0$) the Hubble parameter $H$ in (\ref{gcfe2}) can be expressed in terms of \{$\varphi,\,\dot\varphi$\}. Therefore, in the case $k=0$ the 3-dimensional dynamical system can be reduced to 2-dimensional one, by the analogy with GR \cite{BGZH}. Indeed, since $H$ depends only on scalar field and its derivative, we have the following dynamical system:

\begin{equation}
\left\{
\begin{array}{l}
\dot\varphi=\psi, \\
\dot\psi=-m^2\varphi-\frac{3\psi}{{1-2\beta(\psi^2+m^2\varphi^2)}}\{3\beta m^2\varphi\psi\pm\left(\frac{2\pi}{3M_p^2}\right)^{\frac{1}{2}}[\beta(8m^6\beta\varphi^6+\psi^4-\beta\psi^6)+2\psi^2- \\
\quad\;\;\;\;
-4m^4\beta\varphi^4(2+3\beta\psi^2)+2m^2\varphi^2(1+\beta\psi^2+3\beta^2\psi^4)]^{\frac{1}{2}}\},
\end{array}
\right.
\label{DynSys}
\end{equation}
where $\psi\equiv\dot\varphi$.

However, the expression for $H$ in GTG contains a square root, that could become negative unlike the case of GR (the corresponding square root is always nonnegative). Consequently, phase space is divided into two parts, namely physical region, when scalar field is real and non-physical regions, when scalar field contains imaginary part. Condition

\begin{equation}
\beta(8m^6\beta\varphi^6+\psi^4-\beta\psi^6)+2\psi^2-4m^4\beta\varphi^4(2+3\beta\psi^2)+2m^2\varphi^2(1+\beta\psi^2+3\beta^2\psi^4)=0
\label{Edges}
\end{equation}

determines the edge of the physical regions of the phase space. It is a sixth order algebraic equation and it provides the following three quadratic relations between dynamical variables:

\begin{equation}
\psi^2=\beta^{-1}\left(2\beta m^2\varphi^2-1\right),
\label{Roots}
\end{equation}
\begin{equation}
\psi^2=\beta^{-1}\left(1+2\beta m^2\varphi^2 \pm\sqrt{1+6\beta m^2\varphi^2}\right).
\label{Roots23}
\end{equation}

\begin{figure}[ht]
\begin{center}
\epsfig{file=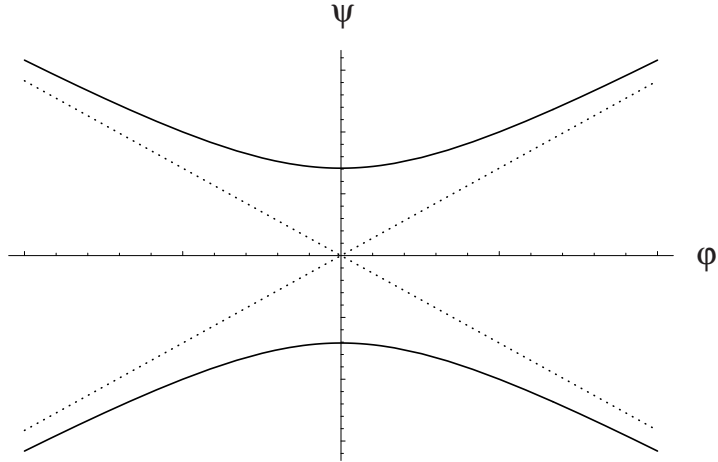,width=4in}
\end{center}
\caption{The edges of the physical phase space. Above and below upper and lower hyperbolas correspondingly dynamical variables contain imaginary parts. The tilt of asymptotes does not depend on the value of $\beta$ and depend only on $m$ (see (\ref{Asympt})).}
\label{fig:edge}
\end{figure}

From now on, we assume $\beta<0$, since in the case of scalar field nonsingular solutions were obtained only for negative $\beta$ \cite{mv1}. In the case $\beta<0$ the edges of the physical region are determined by (\ref{Roots}) only. They are hyperbolas (see fig.\ref{fig:edge}) with asymptotes

\begin{equation}
\psi=\pm\sqrt{2}m\varphi.
\label{Asympt}
\end{equation}

Notice, that asymptotes do not depend on the value of parameter $\beta$.

It is possible to show, that in the case $\beta<0$ sign "+" in (\ref{DynSys}) corresponds to positive Hubble parameter, and, sign "-" takes place for negative Hubble parameter. Therefore for analysis of stage of expansion we will take equations (\ref{DynSys}) with sign "+", and for contraction stage we will use sign "-" in (\ref{DynSys}).

\section{Qualititive analysis}

We will formally consider infinite phase space, i.e. phase trajectories could go to infinity. Notice, however, that from the physical point of view, when the energy density (\ref{phicon}) becomes larger than the planckian value ($\rho\geq M_p^4$), quantum gravity is needed. Therefore clear physical interpretation is possible only for those trajectories, that lay inside the quantum edge $\rho= M_p^4$.

In the finite region of the phase plane \{$\varphi,\,\psi$\} there is only one singular point, namely the point \{$0,\,0$\}. In the limit $\{\varphi\rightarrow 0$,\,$\psi\rightarrow 0\}$ the system (\ref{DynSys}) is reduced to the case of GR. Therefore, all consequences of the analysis of the corresponding dynamical system, that is given in \cite{BGZH}, take place in the system under consideration.

\begin{figure}[ht]
\begin{center}
\epsfig{file=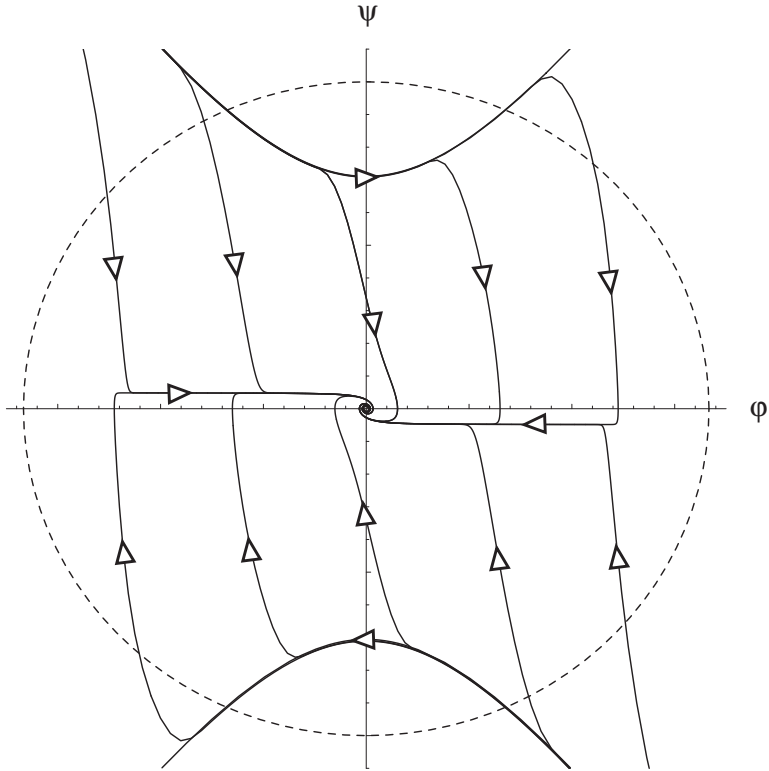,width=4in}
\end{center}
\caption{Phase trajectories of flat cosmological models with massive scalar field within framework of GTG. Dashed line marks the quantum edge (see text). The picture corresponds to expansion stage.}
\label{fig:00}
\end{figure}

Phase trajectories are shown at fig.\ref{fig:00}. Quantum edge (dashed line) marks the distance, at which the energy density is equal to the planckian value. Beyond the quantum edge classical interpretation of solutions becomes impossible. The edge of the physical region could come inside the quantum edge if $|\beta|>M_p^{-4}$ (see (\ref{Roots})).

On the stage of expansion trajectories, starting at some particular point $\{\varphi_0,\,\psi_0\}$ tend to inflationary separatrix, which lays close to the axis $\varphi$. Since at the inflationary separatrix $\psi\ll\varphi$ (and $m\ll M_p$) the effective equation of state is $p\approx-\rho$, i.e. inflationary stage takes place \cite{BGZH}. Therefore, most solutions with particular initial conditions are inflationary. After inflationary stage the field oscillates near the point \{$0,\,0$\} with damped amplitude. All phase trajectories finish at the spiral equilibrium point \{$0,\,0$\}.

The difference between GR and GTG cases is the presence of the edge of the physical region. Initial conditions cannot lay beyond these edges. On the stage of expansion they are repulsive separatrices. 

On the opposite, on the stage of contraction the inflationary separatrix is repulsive, and most initial conditions lead to the effective equation of state $p\approx\rho$. It is well known, that within the framework of GR cosmological models with such equation of state have singularity in future at the stage of contraction. However, since in the GTG case there are two edges of the physical regions, the subsequent evolution of the model changes significantly.

The edges of the physical region become attractive separatrices at the stage of contraction. Most trajectories finish at these edges. Notice, that since the edges are hyperbolas, i.e. infinite curves, dynamical variables tend to infinity in the course of time.

Since the main difference between GR and GTG cases occurs at the contraction stage, below we consider $H<0$ everywhere. Note, that at the stage of contraction fig.\ref{fig:00} must be reflected with respect to the vertical axis.

\section{Phase trajectories at infinity}

The asymptotic behavior can be characterized by the effective equation of state. Taking into account (\ref{phicon}) and (\ref{Asympt}) one arrives to

\begin{equation}
p=\frac{\rho}{3}.
\label{effeqst}
\end{equation}

Thus, within the framework of GTG the effective equation of state for cosmological models with scalar field is the equation of state for ultrarelativistic gas \footnote{The same equation of state appears also for nonlinear scalar field with effective potential $V=\frac{1}{4}\lambda\varphi^4$ and we believe that it is quite a common feature.}. Note, that in the case of equation of state (\ref{effeqst}) the cosmological equations (\ref{gcfe},\ref{Heqn1}) reduce to Einstein equations of GR.

Hubble parameter can be calculated when the scalar field reaches the separatrix (\ref{Roots}). The corresponding value is

\begin{equation}
H=-\frac{m^2\varphi}{\sqrt{2m^2\varphi^2-\beta^{-1}}},
\label{HatHyp}
\end{equation}

and tends to asymptotic value

\begin{equation}
\lim_{\varphi\rightarrow\pm\infty} H=-\frac{m}{\sqrt{2}}.
\label{HatAs}
\end{equation}

Therefore the Hubble parameter becomes constant and relatively small, taking into account that typical values of the mass are $m\sim 10^{-6}M_p$ \cite{Linde}. In infinite future the contraction with constant speed will lead to singularity. Most trajectories finish at this state on the stage of contraction.

To find all singular points of the system (\ref{DynSys}) we use the standard technique of projection of infinite space into the lower hemisphere of the sphere with unit radius, so called Poincare sphere. First of all, we transform cartesian coordinates \{$\varphi,\,\dot\varphi$\} into polar, then project polar coordinates into the hemisphere and, finally, transform back to cartesian coordinates.

The resultant transformations are

\begin{equation}
\left\{
\begin{array}{ll}
u=\frac{1}{\sqrt{1+\left(\frac{\psi}{\varphi}\right)^2}}arctg\left[(\psi^2+\varphi^2)^{\frac{1}{2}}\right], \\
\\
v=\frac{\psi}{\varphi}\frac{1}{\sqrt{1+\left(\frac{\psi}{\varphi}\right)^2}}arctg\left[(\psi^2+\varphi^2)^{\frac{1}{2}}\right].
\end{array}
\right.
\label{Transforms}
\end{equation}

\begin{figure}
\begin{center}
\[ \parbox{3.5in}{
\epsfig{file=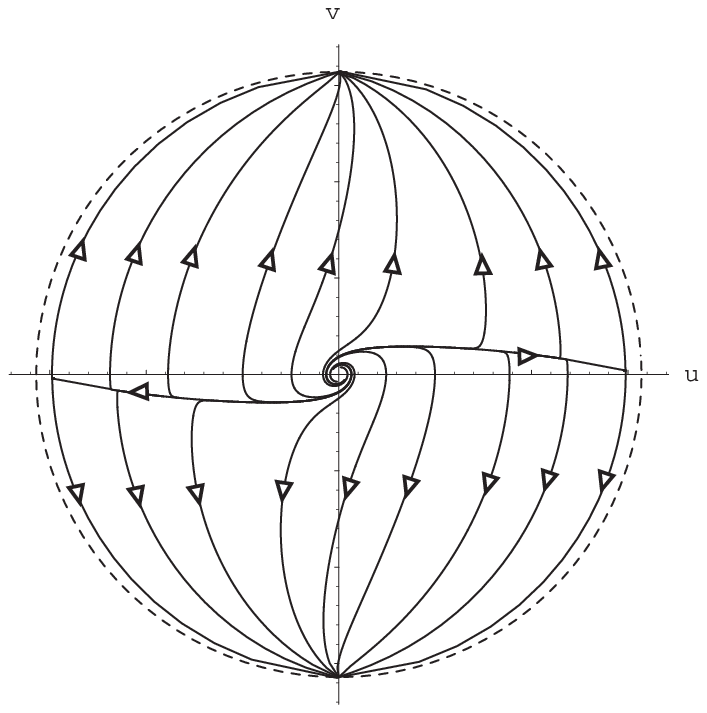,width=3in}}
\parbox[r]{3in}{
\epsfig{file=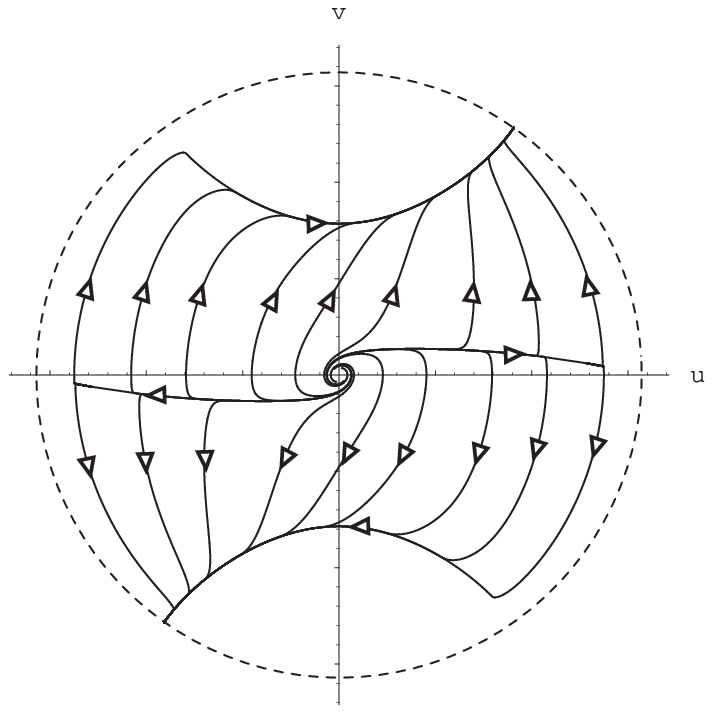,width=3in}} \]
\end{center}
\caption{Phase portraits of the flat cosmological models with massive scalar fields in the cases of GR (left figure, see \cite{BGZH}) and GTG (right figure), projected on the Poincare sphere. The dotted circle corresponds to the infinite values of the dynamical variables. The contraction stage is shown, $H<0$. In order to get phase portrait at the stage of expansion figures have to be reflected with respect to the vertical axis.}
\label{fig:infty}
\end{figure}

The result \footnote{In effect, the radius of the circle, that corresponds to transformations (\ref{Transforms}) is $\pi/2$.} is illustrated at fig.\ref{fig:infty} where left figure corresponds to the GR case, and the right figure corresponds to the case of GTG. The case of GR was investigated at paper \cite{BGZH} and here we provide the corresponding figure for comparison.

There are two knots, two saddles and one spiral focus (that we discussed in the previous section) in both cases. In the GR case knots lay at the axis $u=0$ and saddles lay at the axis $v=0$ at infinity of the real phase space.

In the case of GTG two knots lay at the point, where asymptotes (\ref{Asympt}) cross the radius of the circle. Notice again, that the tilt of asymptotes does not depend on parameter $\beta$. Moreover, in new variables the form of equation (\ref{Asympt}) is conserved.

It could be seen, that the tilt of asymptotes (\ref{Asympt}) tends to zero with zero mass, i.e. it is possible to make the hyperbolas (\ref{Roots}) as flat as we wish. Taking into account that at the stage of contraction the hyperbolas (\ref{Roots}) are attractive separatrices, the question arise, what is the effective equation of state of the scalar field on these hyperbolas.

From (\ref{phicon}) and (\ref{Roots}) we have

\begin{equation}
p=\rho\left(\frac{1}{3}+\frac{2}{3(1-3\beta m^2\varphi^2)}\right).
\label{EQST}
\end{equation}

Therefore

\begin{equation}
\frac{\rho}{3}\leq p\leq \rho.
\label{Lims}
\end{equation}

The two limiting cases take place when the scalar field and its derivative with respect to time tend to infinity and to zero correspondingly.

\section{Discussion}

It was argued in \cite{Min01},\cite{Min03} that it is possible to built nonsingular inflationary cosmological models within the framework of GR. However, in spite of the fact that such models could exist, according to the analysis of \cite{BGZH} their measure is equal to zero, because on contraction stage effective equation of state tends to $p=\rho$, that lead to cosmological singularity for most solutions.

According to \cite{Min03}, the possibility to avoid cosmological singularity remains if "physical conditions leading to the bounce take place at the end of cosmological compression". It should be clear, however, that in this case nonsingular solutions have much less attractive character, because they necessarily require special physical conditions,  for instance phase transitions, at the bounce. Such models are already discussed in the literature \cite{BP},\cite{IR}.

The analysis presented in the paper shows that the similar situation takes place within the framework of GTG in the case of flat models. Since cosmological equations of GTG are much more complicated than than of GR, they possess a wide spectrum of solutions that is absent in the case of GR. In particular, there are oscillating solutions \cite{Ver03} in the case of closed models for the Universe, filled by nonlinear scalar field. However, such solutions have nothing to do with inflation.

We conclude therefore that attempts to build nonsingular inflationary cosmological models within framework of GR as well as GTG lead to serious complications connected with domination of scalar field on the stage of contraction.

\section{Conclusions}

The analysis carried out in this paper shows that the dynamics of the flat cosmological models with massive scalar field within the framework of GTG differs significantly from the corresponding behavior within GR. In the phase space there are regions when the scalar field acquires imaginary part. Corresponding regions are unphysical. The edges of the physical regions are hyperbolas,  symmetric with respect to the axis $\dot\varphi$.

At the stage of expansion most phase trajectories, that start at some point inside the quantum edge, i.e. the region, where the energy density is smaller than the planckian value, tend to inflationary separatrix. Therefore, inflationary solutions are quite generic feature of cosmological models within GTG.

At the stage of contraction the dynamics of cosmological models within framework of GTG changes significantly in comparison with corresponding dynamics within GR due to the presence of attractive separatrices. Before reaching the separatrix the effective equation of state of the model is $p\approx\rho$ like in the case of GR. On the separatrix the equation of state tends to the limiting case $p=\rho/3$.

The Hubble parameter at the stage of compression tends to the limiting value $H=-\frac{m}{\sqrt{2}}$, and corresponding models have a cosmological singularity in infinite future. Therefore, nonsingular inflationary solutions are unstable within the framework of GTG. This conclusion is applicable in part to closed and open cosmological models with negligible curvature due to continuity of the phase space.

\end{document}